# Enhanced diffusion and enzyme dissociation

Ah-Young Jee†, Kuo Chen‡, §, Tsvi Tlusty†, ∥, Jiang Zhao‡, §, and Steve Granick*†, ∥, ¶

†Center for Soft and Living Matter, Institute for Basic Science (IBS), Ulsan 44919, South Korea

‡Beijing National Research Center for Molecular Sciences, Institute of Chemistry, Chinese, Academy of Sciences, Beijing 100190, China

§University of Chinese Academy of Sciences, Beijing 100049, China

∥Department of Physics, UNIST, Ulsan 44919, South Korea

¶Department of Chemistry, UNIST, Ulsan 44919, South Korea

**ABSTRACT:** The concept that catalytic enzymes can act as molecular machines transducing chemical activity into motion has conceptual and experimental support, but much of the claimed support comes from experimental conditions where the substrate concentration is higher than biologically relevant and accordingly exceeds $k_M$, the Michaelis-Menten constant. Moreover, many of the enzymes studied experimentally to date are oligomeric. Urease, a hexamer of subunits, has been considered to be the gold standard demonstrating enhanced diffusion. Here we show that urease and certain other oligomeric enzymes of high catalytic activity above $k_M$ dissociate into their smaller subunit fragments that diffuse more rapidly, thus providing a simple physical mechanism of enhanced diffusion in this regime of concentrations. Mindful that this conclusion may be controversial, our findings are supported by four independent analytical techniques, static light scattering, dynamic light scattering (DLS), size-exclusion chromatography (SEC), and fluorescence correlation spectroscopy (FCS). Data for urease are presented in the main text and the conclusion is validated for hexokinase and acetylcholinesterase with data presented in supplementary information. For substrate concentration regimes below $k_M$ at which these enzymes do not dissociate, our findings from both FCS and DLS validate that enzymatic catalysis does lead to the enhanced diffusion phenomenon.

## INTRODUCTION

The ubiquity of enzyme catalysis in biology and technology has become even more interesting with the discovery that enzyme catalysis appears to transduce chemical activity into motion leading to enhanced diffusion, a conclusion that came originally from experiments[1-9] and now is buttressed by theoretical analysis[7-8, 10-16]. Much of the experimental support comes from considering enzymes of high catalytic turnover, among them urease, acetylcholinesterase and hexokinase, the three enzymes that we consider in this study. We are motivated by noticing that these enzymes are oligomeric and evolved to operate within biological cells at substrate concentrations below the Michaelis-Menten constant $k_M$ which for urease is ≈3 mM.[17] As many (not all) of the experiments demonstrating enhanced diffusion operate at significantly larger substrate concentrations, it is interesting and relevant to inquire into origins of enhanced diffusion when the substrate concentrations exceed those that are biologically relevant. We focus on urease, which has been considered to be the gold standard demonstrating enhanced diffusion[1-4, 7-9, 14, 18]. The product of urease catalysis is gas whose presence might influence mobility, $CO_2$. For generality, we also study other enzymes, hexokinase, and acetylcholinesterase.

Fluorescence-based measurements of diffusion in the urease system show that it grows in two steps. This enzyme's effective diffusion coefficient measured by fluorescence correlation spectroscopy (FCS) grows smoothly with increasing substrate concentration up to $k_M$ and saturates at a plateau of ≈25% enhancement.[8] We interpreted this regime in terms of enzyme leaps stimulated by the catalytic activity such that chemical activity led to the enhanced mobility.[8] When the substrate concentration was further increased a second rise of enhanced diffusion was observed, up to 80%[8] faster than in the absence of substrate. The observation of two-step rise is intriguing because the second concentration regime, substrate concentrations in the 0.1-1 M regime,[1-4, 14, 18] was the condition of many prior experimental studies. We speculated that the second regime of extra-enhanced diffusion might reflect enzyme dissociation[8] but made no direct test of this hypothesis for this system, though enzyme dissociation into subunits was reported already long ago for $F_1$-ATPase[19] and more recently, discussed for other oligomeric enzymes.[20-21]

Meanwhile, concerns were raised that fluorescence-based measurements might introduce experimental artifact incorrectly interpreted as enhanced diffusion.[9, 20-23] With these considerations in mind, here we revisit the urease system and test the enzyme dissociation hypothesis. Mindful that our conclusions may be controversial, our conclusions are tested by four independent analytical techniques: static light scattering, dynamic light scattering (DLS), size-exclusion chromatography (SEC), in addition to fluorescence correlation spectroscopy (FCS). Buffer conditions and other experimental protocols are specified in Supplementary Material.

## RESULTS AND DISCUSSION

**Choice of enzyme samples.** To the best of our knowledge, all studies of enhanced diffusion in the urease system concern urease extracted from *Canavalia ensiformis*, the common Jack bean. The source of the urease sample was specified in some studies [1-2, 8], not specified in other studies [3-4, 7, 9, 14, 18], as summarized in the Table S1 (supplementary information). To anticipate conclusions of the following discussion, we found it reassuring that despite quantitative differences according to which source of urease we used, the qualitative conclusions were the same.

Bearing in mind the doubts recently expressed whether FCS is a true measure of translational diffusion [9, 20-23], we were motivated to perform experiments independent of and complementary to FCS. In order to make the findings most comparable to FCS (fluorescence) measurements on which relied so much earlier data in the literature [1-4], our principal independent tests were performed on enzymes also labeled in the same manner as for FCS experiments with fluorescent dye using the procedures described in Supplementary Material. Specifically, the light scattering experiments were performed on urease labeled with fluorescent dye.

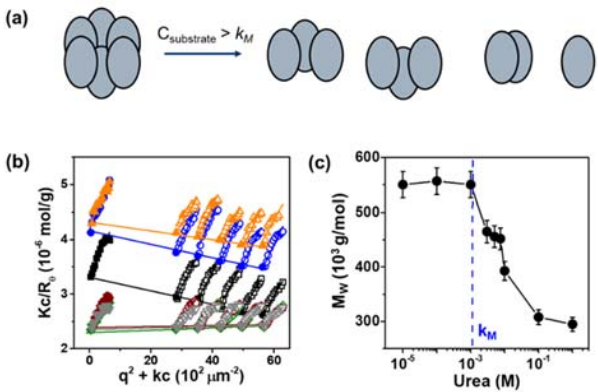

Figure 1. Static light scattering of urease. (a) Schematic diagram in which a multimeric enzyme may dissociate into subunits. (b) Zimm plot for sample $Ur_{1f}$ at various urea concentration where c is mass concentration of enzyme, q is wavevector, and the symbols K, R and k are constants with standard textbook meanings in static light scattering. K is optical constants, R is the Rayleigh ratio, and k is a constant chosen arbitrarily to shift curves on the x-axis according to the Zimm Plot method. Data are open symbols, plotted from top towards bottom at progressively smaller c. Filled symbols denote these data extrapolated to zero concentration. Lines are least squares fits to the data. Yellow, blue, black, brown, grey, and green shows urea concentration 1 M, $10^{-1}$ M, $10^{-2}$ M, $10^{-3}$ M, $10^{-4}$ M and $10^{-5}$ M in 100 mM PBS buffer (pH 7.2), respectively. (c) Weight-average molecular weight of urease, which is the inverse of the y-intercept in (b), plotted against urea concentration.

**Table I. Enzymes studied and their Michaelis-Menten characterization.** Characterization was done in this laboratory except when identified by literature reference.

| Code | Sample | Dye | $k_{cat}$ (s$^{-1}$) | $k_M$ (mM) |
|---|---|---|---|---|
| $Ur_{1u}$ | Low activity urease (type IX) | unlabeled | 3,040 | 1.04 |
| $Ur_{1f}$ | Low activity urease (type IX) | labeled | 2,140 | 1.08 |
| $Ur_{2u}$ | High activity urease (type C-3) | unlabeled | 45,020 | 2.60 |
| $Ur_{2f}$ | High activity urease (type C-3) | labeled | 17,020 | 1.20 |
| $Ac_u$ | Acetylcholinesterase (type VI-S) | unlabeled | 6,100 | 0.50 |
| $Ac_f$ | Acetylcholinesterase (type VI-S) | labeled | 4,500 | 0.52 |
| Hex | Hexokinase | unlabeled | 250 | 0.20 |

We now mention some differences between the samples listed in Table S1, especially our finding that the urease of highest catalytic activity aggregated when its concentration exceeded nM. Indeed, the tendency of urease to aggregate in 100 mM PBS buffer (pH 7.2) when the enzyme concentration exceeded 100 nM was noted by us earlier.[7-8] Those experiments were performed using enzymes with the highest purity available to us commercially, Sigma-Aldrich "Type C-3 urease." This presented a difficulty as we wished to respond constructively to the voiced concerns that FCS is artifact [9, 20-23], yet only FCS possesses the sensitivity needed to measure diffusion at nM concentrations. Therefore to assess this sample with our complementary experiments was not feasible.

Of all the experiments one might use to test the conditions under which oligomeric enzymes might dissociate (Fig. 1a), scattering experiments are the most direct as they can give absolute measurements of molar mass. Using the sample of highest catalytic activity, we attempted static light scattering at nM conditions where FCS showed the absence of aggregation, but they failed owing to insufficient sensitivity. Therefore, we turned to using a sample that we found to be less aggregation-prone, Sigma-Aldrich "Type IX". In what follows, we refer to this as sample $Ur_1$, and to the sample of higher catalytic activity used in our earlier experiments [7-8] as sample $Ur_2$.

Although control experiments showed the same qualitative conclusions regardless of labeling (Supplementary Information), we found that labeling the enzyme with fluorescent dye modulated the catalytic activity, probably by modulating access to the active site. In what follows, we refer to unlabeled and dye-labeled urease as samples $Ur_{1u}$ and $Ur_{1f}$, respectively. Labeling and purification processes are described in Supplementary Information.

Table I summarizes the three enzymes studied (urease, acetylcholinesterase, hexokinase). For each enzyme, we determined the Michaelis-Menten constants: turnover rate, $k_{cat}$, and the Michaelis-Menten constant, $k_M$. Sources and experimental procedures are described in Supplementary Material. Fig. S1 shows Michaelis-Menten kinetic curves for Samples $Ur_{1f}$ (urease) and $Ac_f$ (acetylcholinesterase).

**Static light scattering.** First, the absolute weight-average molecular weight ($M_w$) of urease sample $Ur_{1u}$ was determined using static light scattering.[35] The so-called Zimm plot is the standard way to analyze data of this kind. On the ordinate, a quantity proportional to sample concentration (c) is multiplied by instrumental constants (K) and divided by a measure of scattering at a given specified angle ($R_\theta$). On the abscissa, one plots wavevector squared ($q^2$) shifted by concentration (kc) according to the standard method of fitting, the so-called Zimm plot. Extrapolating both wavevector and concentration to zero, one obtains the y-intercept, which is the inverse weight-average molecular weight, $M_w$. At substrate concentrations below $k_M$ this gave $M_w = 5.5 \times 10^5$ g-mol$^{-1}$ (Da), consistent with the known hexameric form of this enzyme.[17]

The substrate concentration was then increased in small increments by up to 4 orders of magnitude, up to 1 M. It is obvious in Fig. 1b that $M_w$ decreases. Inspecting a plot of $M_w$ against substrate concentration (Fig. 1c), one sees that $M_w$ is constant below 1 mM but decreases when the substrate concentration is higher. At 1 M concentration the molecular weight is slightly above one-half the original value, suggesting that in the presence of urea, this enzyme became heavily dissociated. Dissociation into trimers was not complete as $M_w$ slightly exceeded one-half the initial value.

Slopes of Zimm plots quantify pairwise interactions as they are proportional to the second virial coefficient, $A_2$; positive and negative slopes imply repulsion and attraction, respectively. The negative $A_2$ at substrate concentrations above 10 mM, more strongly so with increasing substrate concentration, indicates that pairwise attraction grows with increasing concentration (Fig. S2), indicating growing tendency towards aggregation. Control FCS measurements described below confirm the same pattern of two-regime enhanced diffusion, below $k_M$ and above it, as reported earlier for Sample $Ur_{2f}$.[8]

**Dynamic light scattering (DLS).** The static light scattering experiments measured molar mass, not diffusion. In order to measure diffusion without using FCS, we turned to dynamic light scattering (DLS). This standard method quantifies the photon autocorrelation function and extracts from it the implied translational diffusion coefficient $D$. From this, one infers the hydrodynamic radius $R_h$ of an equivalent sphere.[37]

Using sample $Ur_{1f}$, measurements were made for a relatively short time, as soon as feasible to do after adding substrate (30 s), to minimize the opportunity for aggregation. In the absence of substrate and under conditions of very low substrate concentration, the measured $R_h \approx 8.5$ nm is consistent with literature values for the radius of urease.[36]

Fig. 2a compares the autocorrelation G(t) below and above $k_M$ that we determined for this sample (Fig. S1). The curve for the latter is shifted to faster time lags indicating faster diffusion, and also shows a two-step process, obvious to the eye in this curve. This contributes to a bimodal distribution when these curves are deconvoluted to show the relative abundance of diffusing entities of different hydrodynamic radius $R_h$ as plotted in Fig. 2b. To perform convolution we used the standard CONTIN algorithm.[38] The bimodal distribution at high substrate concentration shows one peak close to the original one, and also a second peak of the size expected if urea dissociates into trimers.

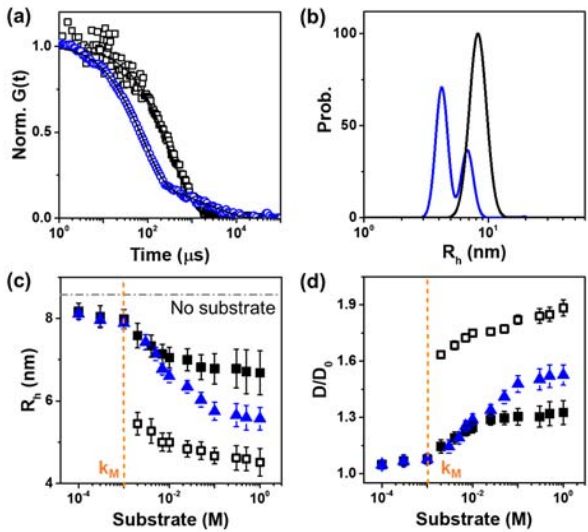

Figure 2. Dynamic light scattering of urease sample $Ur_{1f}$. (a) Photon autocorrelation function is plotted against time lag for a representative substrate concentration below $k_M$ (0.1 mM, black) and a representative substrate concentration above $k_M$ (1M, blue). (b) Distributions of hydrodynamic radius $R_h$ inferred using the CONTIN algorithm from data in panel a. Relative abundance is plotted against radius. The $R_h$ of black peak is consistent with the reported value.[36] (c) Hydrodynamic radius $R_h$ is plotted against logarithmic substrate concentration across 4 orders of magnitude. Low-$R_h$ peak of the bimodal distribution (empty black), high-$R_h$ peak (filled black) and average $R_h$ weighted by relative abundance (blue) are shown. (d) Relative diffusion coefficients implied from data in panel c are plotted against logarithmic substrate concentration across 4 orders of magnitude. Symbols are same as in panel c.

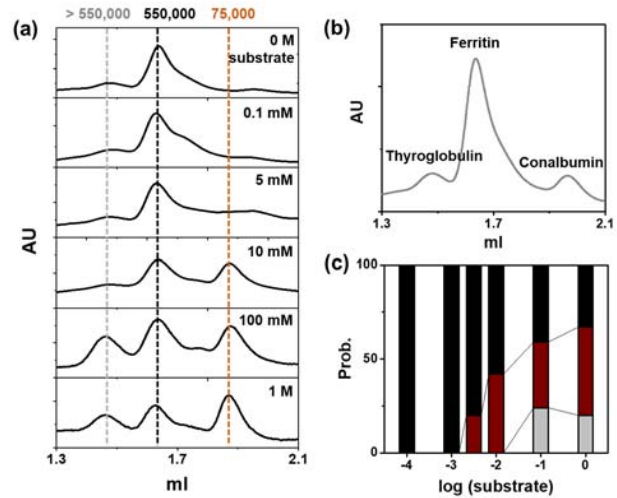

Figure 3. Size-exclusion chromatography of urease sample $Ur_{1u}$ in 100 mM PBS buffer (pH 7.2). (a) plot of chromatograms on a Superose 6 column. Relative volume eluted through the column is plotted for 6 substrate concentrations, 0 M, 0.1 mM, 5 mM, 10 mM, 100 mM, and 1 M urea, after calibrating the column with standard proteins. (b). (b) Calibration of the SEC column. Elution profiles of Calibration Kit proteins in 100 mM

PBS buffer (pH 7.2) (standard proteins) on Superpose 6 column. Elution volumes ($V_e$) are identified with maximum peak height of each respective protein. Thyroglobulin, ferritin, and conalbumin have molar mass 669,000 g/mol, 440,000 g/mol, and 75,000 g/mol, respectively. (c) Relative sizes of the eluted urease, extracted from the peaks of each chromatogram, are plotted against logarithmic substrate concentration. The ordinate of this bar graph shows the relative abundance of the hexamer (black), trimer (red) and dimer (grey). The relatively high enzyme concentration needed for this experiment is believed to explain quantitative differences between Fig. 3c and Fig. 4d.

From these distributions we took the peak maxima, calculated their abundance-weighted averages, and plotted these quantities against substrate concentration in Fig. 2c. Finally, diffusion coefficients were calculated from the Stokes-Einstein equation. Diffusion enhancement of this enzyme relative to the substrate-free situation is plotted in Fig. 2d against substrate concentration. Note the peak from highest $R_h$, which diminishes with increasing substrate concentration, the peak with lower $R_h$, which also diminishes with increasing substrate concentration, and the average inferred from the average $R_h$.

**Size-exclusion chromatography (SEC).** This standard method to characterize enzyme purity[39] was implemented by us by measuring elution through a Superose 6 SEC column (GE Healthcare), which has a measurement range from 5,000 to 5,000,000 Da. The column was calibrated using standard proteins (thyroglobulin, ferritin, conalbumin), as shown in Figure 3b. This allowed the approximate molecular weight of individual peaks of our unknown sample to be determined.

For Sample $Ur_{1f}$, representative elution curves are plotted in Fig. 3a. In the absence of substrate, the SEC chromatogram of urease shows one major peak at elution volume $V_e$ = 1.6 ml, and from comparison to the peptide standards this corresponds to 550,000 g-mol$^{-1}$, the molecular weight of urease hexamer. From 5 mM urea and above, a slight shoulder appears on the higher elution side, indicating generation of smaller units. With increasing urea, this becomes a distinct peak centered at 75,000 g-mol$^{-1}$. There are also signs of aggregation. For 100 mM urea, but not yet for 10 mM urea, a second distinct peak appears at $V_e$ = 1.5 ml, and this is assigned to 700,000 g-mol$^{-1}$, some kind of aggregate that grew with further increase of urea concentration. Focusing on dissociation of the enzyme into subunits, we deconvoluted the elution peak areas to give relative abundance of hexamers, trimers, and dimers as a function of substrate concentrate, as plotted in Figure 3c.

The time to make measurements using SEC is at least one hour to elute after the sample solution is injected into the column. Unlike the measurements we made using static and dynamic light scattering, which were completed within a few minutes, the SEC experiment therefore was more sensitive to the slow process of protein aggregation. The relatively high enzyme concentration needed for this experiment, 200 nM (Supplementary Information), is believed to explain quantitative differences between Fig. 3c and Fig. 4d. Aggregation is suspected to involve denatured urease but as aggregation was not the point of this study, this matter was not pursued.

**Intensity-weighted FCS.** Fluorescence correlation spectroscopy (FCS), a standard method to measure the diffusion of nM-level quantities of molecules including proteins, was used by us and others in earlier studies of enhanced diffusion. The measured autocorrelation curves G(t) nicely fit a free diffusion fitting model regardless of urea concentration, except that upon inspecting the fitting residuals for high urea concentration, small systematic deviations are observed at the most rapid time scales, faster than hundreds of microseconds (Fig. 4a). As this time scale is a minor contribution to the overall fit, one-component fitting was used. Customarily analyzed from the intensity-intensity autocorrelation function, raw data from this experiment consists of fluorescence intensity traces as a function of time. In fact, perfect dye labeling efficiency is impossible, but the labeling protocol uses an excess of dye, so for this argument we assume that the dye has labelled all subunits. To the extent this argument holds, it is therefore relevant to consider how raw values of the fluorescence intensity may change.

The intensity was time-independent during measurements in buffer and 1 mM urea. On the other hand, the intensity gradually diminished over time in the presence of 1 M urea (Fig. 4b). Pursuing these differences and using sample $Ur_{1f}$ in the absence of substrate and at substrate concentrations below $k_M$, we observed a nearly-Gaussian intensity distribution. For higher substrate concentrations this became progressively broader, so we deconvoluted the intensity distributions as illustrated in Fig. 4c. The idea behind deconvolution was that if unperturbed urease hexamer enzymes are uniformly labeled to display intensity $I_{max}$ when passing through the FCS confocal spot, dissociated trimers will display $(1/2)I_{max}$ and dimers will display $(1/3)I_{max}$. Deconvolution was performed according to this reasoning. As a function of substrate concentration, the fluorescence intensity was separated into relative abundance of hexamers, trimers, and dimers, as plotted in Figure 4d.

A technical point is that in this analysis, certain quantitative differences can result depending on how the intensity traces are binned according to time. Single Gaussian and bimodal distributions are a robust conclusion regardless of binning size at 1 mM and 10 mM urea, which are conditions where urease has experienced little dissociation into subunits. On the other hand, when considering the 100 mM substrate concentration, 1 ms binning caused self-averaging of short-time events. Under this condition, we found that binning at 0.2 ms revealed a trimodal rather than the bimodal distribution implied by 1 ms binning. These considerations are believed to be why intensity traces do not show evident of the smallest oligomeric subunits, monomers and dimers (Fig. S3).

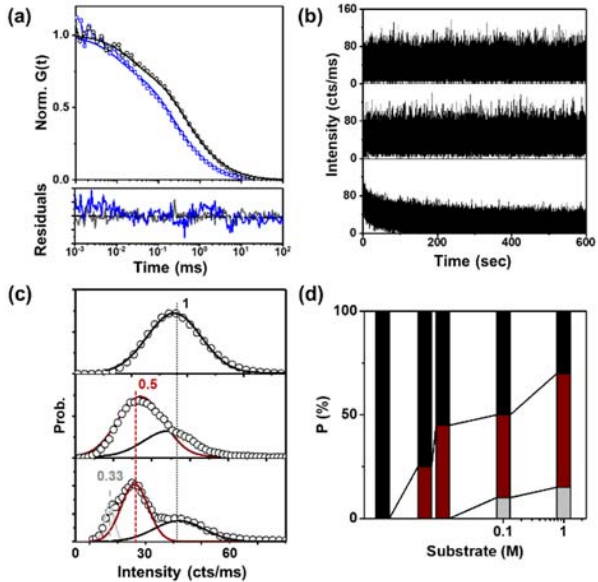

Figure 4. FCS experiments. (a) Normalized autocorrelation function G(t) of $Ur_{1f}$ in 1 mM urea (black) and 100 mM urea (blue) in PBS buffer. The bottom panel shows fitted residuals. (b) Fluorescence intensity time trace of urease with no substrate present; 1 mM urea; and 1 M urea, from top panel to bottom. (c) Fluorescence intensity distribution of urease sample $Ur_{1u}$ at different urea concentration regime. From top panel to bottom panel, it shows 1 mM, 10 mM, 100 mM of urea, respectively. Black, red, and grey fitting curves represent $I_{max}$, $1/2 I_{max}$, $1/3 I_{max}$ respectively. Binning time, 0.2 ms. (d) The dissociation from hexameric urease to trimer and dimer according to substrate concentration obtained from area fraction of each distribution. Black, red, and grey shows hexamer, trimer, and dimer, respectively.

**Another oligomeric enzyme, acetylcholinesterase.** Bearing in mind that urea is a common protein denaturation agent,[40-41] a fact that potentially might influence the action of urea on the enzyme urease despite the fact that urease has been considered a model system in which to study enhanced diffusion, generality of these findings was checked regarding acetylcholinesterase (AChE), another oligomeric enzyme that in the literature was interpreted to display enhanced diffusion at substrate concentrations above $k_M$.[7] AChE is a tetramer and its substrate is acetylcholine. We denote the unlabeled and dye-labeled samples as $Ac_u$ and $Ac_f$, respectively. Studying sample $Ac_f$, this enzyme's hydrodynamic radius was measured by DLS and the distributions of $R_h$ were inferred when the substrate concentration was increased to values well above $k_M$. As shown in Fig. S4, these data follow the same dissociation patterns as urease.

**Another oligomeric enzyme, hexokinase.** Hexokinase I, a dimeric enzyme of size 104,000 g/mol used in several earlier studies for which enhanced diffusion was reported at substrate concentrations above $k_M$ was also investigated.[5] The substrate is glucose. This enzyme's hydrodynamic radius was measured by DLS and the distributions of $R_h$ were inferred when the substrate concentration was increased to values well above $k_M$. The data are similar to those presented above for urease and acetylcholinesterase (Fig. S5). Regarding enhanced diffusion, the data obtained by FCS at high substrate concentrations are intermediate between the $D/D_0$ of the undissociated enzyme and its dissociated components, as expected of this measurement that does not distinguish between them.

**Comparing enzymes with different commercial provenance.** To assess generality, the remaining samples in Table I were also studied for completeness. For the additional samples, their Michaelis-Menten characterization is shown in Fig. S6.

For urease, DLS experiments are compared for samples $Ur_{1u}$ (Fig. S7) and $Ur_{2f}$ (Fig. S8). For each, the enzyme's hydrodynamic radius was measured by DLS and the distributions of $R_h$ were inferred when the substrate concentration was increased to values well above $k_M$. Between the samples there is excellent consistency with quantitative differences. These may reflect differences of turnover rate.

For acetylcholinesterase, similar comparisons were made for a sample unlabeled with fluorescent dye, sample $Ac_u$ (Fig. S9). Between the samples there is excellent consistency.

**Static and dynamic measurements compared.** We were interested to compare diffusion from different experiments. To do this, it was reasonable to suppose that hydrodynamic radius $R_h$ equals static radius of gyration $R_g$ within our experimental uncertainty. However, $R_g$ measured by static light scattering is notorious for having high experimental uncertainty in the regime of our relatively-low molar mass, so instead we estimated $R_g$ from the measured $M_W$. Our reasoning was to identify $R_g$ with the radius of the equivalent sphere, knowing that globular proteins have density ≈1 g-cm$^{-3}$. From $R_g \approx R_h$, $D$ was calculated using the Stokes-Einstein equation.

Fig. 5 compares the diffusion coefficients $D$ inferred for urease from static light scattering, dynamic light scattering, and FCS. From these three independent techniques, on the same scale all the $D$ are plotted against logarithmic substrate concentration, over 5 orders of magnitude of substrate concentration in Fig. 5a, and emphasizing substrate concentrations below $k_M$ in Fig. 5b. All measurements appear to agree when the substrate concentration exceeds $k_M$. For substrate concentrations below $k_M$, the regime in which the enzyme does not dissociate, FCS and dynamic light scattering agree in showing enhanced diffusion.

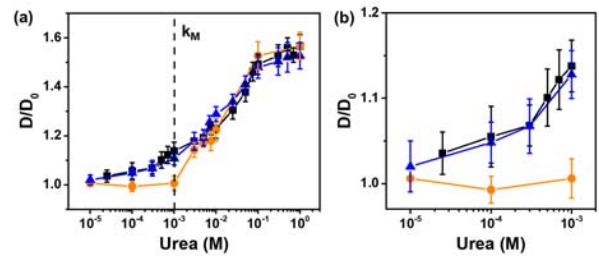

Figure 5. Comparison of static (SLS) and dynamic (DLS and FCS) measurements for urease sample $Ur_{1f}$. Black, blue, and orange show findings from FCS, DLS, and SLS, respectively. Diffusion is estimated from static light scattering using the Stokes-Einstein equation with the approximation $R_g \approx R_h$.

## CONCLUSIONS

Our experiments offer an alternative explanation for many experiments in the literature that were performed at substrate concentrations above $k_M$, as in this regime we confirm that dissociation of oligomeric enzymes into subunits can explain those findings though they do not exclude enhanced that diffusion may contribute. It is interesting to speculate about the biological function of this enzyme dissociation phenomenon. Not known presently is whether this question has functional significance, as such high concentrations are not believed to occur in natural settings. It might function as a biological regulatory mechanism.

At the same time, in the regime of biologically-relevant substrate concentrations below $k_M$, we do observe that the presence of substrate enhances diffusion even of the oligomeric enzyme. This is broadly consistent with the qualitative conclusion from much previous work and helps to clarify the regime of their potential validity.[1-16] In particular, while enhanced diffusion concept needs qualification may not apply in the substrate concentration regime where enzymes dissociate into subunits, they may apply at lesser concentrations, and this is interesting because the regime of lesser concentration is more relevant biologically.

This study is not believed to be directly relevant to an interesting parallel family of studies in which catalytically-active enzymes, urease in many instances[1-4, 7-9, 14, 18, 24-30], were attached chemically to the surfaces of colloidal beads or nanoparticles. Enhanced mobility or ballistic motion of colloidal beads is observed when substrate is added[4, 24-30]. It is unknown how the methods of enzyme surface-attachment might influence the opportunities for enzyme dissociation into subunits, however. Also enzyme-driven colloids are surely influenced by diffusiophoresis produced by a concentration gradient of reaction products near the surfaces of colloidal beads[31-34]. Diffusiophoresis is not believed to contribute to the situations, considered here, of enzymes at nM concentrations.

## ASSOCIATED CONTENT

**Supporting Information**. Additional data related to this paper are present in the Supplementary Materials. Experimental detail, enzyme assays, and more DLS results of various samples.

## AUTHOR INFORMATION

### Corresponding Author
*sgranick@ibs.re.kr

### Author Contributions
The manuscript was written through contributions of all authors. / All authors have given approval to the final version of the manuscript.

### Notes
The authors declare no competing financial interests.

## ACKNOWLEDGMENT
This work was supported by the taxpayers of South Korea through the Institute for Basic Science, project code IBS-R020-D1. For instrument access we thank IBS-R0190D for dynamic light scattering and IBS-R022-D1 for size exclusion chromatography. We are indebted to Dr. Hyun Suk Kim in the IBS Center for Genomic Integrity for help with SEC measurements.

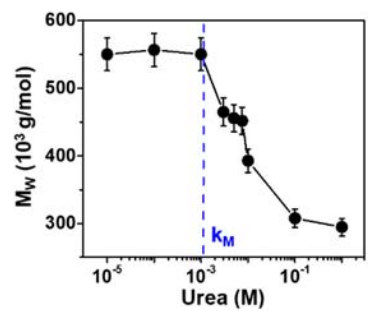

Supporting Information for

# Enhanced diffusion and enzyme dissociation

Ah-Young Jee, Kuo Chen, Tsvi Tlusty, Jiang Zhao, and Steve Granick

Contents

**Experimental procedures**
**Figure S1.** Assays of enzyme activity and fit to the Michaelis-Menten equation for urease sample $Ur_{1f}$ and acetylcholinesterase sample $Ac_f$.
**Figure S2.** Second virial coefficient ($A_2$) as a function of substrate (urea) concentration, determined from static light scattering on urease sample $Ur_{1u}$.
**Figure S3.** Fluorescence intensity distribution of urease sample $Ur_{1u}$ at 100 mM urea with 0.2 ms (upper panel) and 1 ms binning size (lower panel)..
**Figure S4.** Dynamic light scattering of dye-labeled acetylcholinesterase, sample $Ac_f$.
**Figure S5.** Dynamic light scattering of hexokinase, sample Hex.
**Figure S6.** Assays of enzyme activity and fit to the linearized Michaelis-Menten equation for samples not shown in Fig. S1.
**Figure S7.** Dynamic light scattering of unlabeled urease, sample $Ur_{1u}$.
**Figure S8.** Dynamic light scattering of dye-labeled high-activity urease, sample $Ur_{2f}$.
**Figure S9.** Dynamic light scattering of unlabeled acetylcholinesterase, sample $Ac_u$.
**Table S1.** Published literature on urease enhanced diffusion showing the reported source of urease in each study.

**Experimental procedures**

Samples. Urease (type IX, type C-3) from jack bean, purchased from Sigma, was labeled at the amine residue with dylight 488 maleimide dye by a protocol involving 150 mM phosphate buffer (pH 7.2) with added 2 μM urease and 40 μM fluorescent dye solution, stirred for 6 h at room temperature. Acetylcholinesterase from Electrophorus electricus (electric eel), purchased from Sigma-Aldrich, was labeled at its carboxyl residue by Dylight 488-NHS (N-hydroxysuccinimide) dye by a protocol in which 30 μM dye solution and 1 μM enzyme were added to a mixture of 80% phosphate buffer solution (PBS) and 20% dimethyl sulfoxide (DMSO) before 6 h of stirring at room temperature. Finally, the dye-labeled enzymes were purified by removing the free dye by membrane dialysis (Amicon ultra-4 centrifugal filter; Millipore). Hexokinase I from saccharomyces cerevisiae was purchased from Sigma-Aldrich and it was labeled with Alexa fluor 488 labeling kit (Invitrogen) using a protein fluorescence labeling kit (Invitrogen).

Enzyme activity assay. The urease and acetylcholinesterase assays were performed using the urease activity kit (MAK120, Sigma Aldrich) and acetylcholinesterase activity kit (MAK119, Sigma Aldrich) as reported in the manufacturer's instructions. Hexokinase activity listed in Table I was taken from the literature.[1]

Static light scattering (SLS). A commercial laser light scattering instrument (ALV/DLS/SLS-5022F) equipped with a multi-τ digital time correlator (ALV5000) and a cylindrical 22 mW He-Ne laser ($\lambda_0$ = 632.8 nm, Uniphase) was used in the laboratory of Jiang Zhao at the Chinese Academy of Sciences. The measurements were conducted at scattering angles from 30° to 150° in steps of 10°. For SLS measurements, dye-labeled urease and its substrate solutions were mixed at the desired concentration in 100 mM PBS buffer (pH 7.2) and filtered twice by using 100 nm pore size syringe filter (Whatman). The range of urease concentration was 50 nM to 150 nM.

Dynamic light scattering (DLS). A Brookhaven ZetaPALS instrument with the ZetaPlus option at 90° scattering angle was used in the IBS Center for Multidimensional Carbon Materials. For DLS measurements, 30 nM dye-labeled enzymes (Urease, AChE) and the substrate solution (urea for urease, acetylthiocholine for AChE) were mixed at the desired concentration in 100 mM PBS buffer (pH 7.2) and filtered twice using 100 nm pore size syringe filter (Whatman). For hexokinase reaction, 50 nM of dye-labeled hexokinase I was added to the medium containingd 50 mM Tris:HCl pH 7.5, 0.5 mM MgCl$_2$, 0.12 mM ATP, 0.1 mM NAD(P)+, and various concentrations of glucose were mixed and filtered twice using 100 nm pore size syringe filter.

Size exclusion chromatograph. The SEC measurements were carried out at room temperature, using Superose 6 column (GE Healthcare) with an AKTA Explorer FPLC system (GE Healthcare). For molecular weight detection, urease mixed with its substrate at the desired concentration in 100 mM PBS buffer (pH 7.2) were injected into the column at an eluent flow rate of 0.2 mL/min and the absorbance was measured at 280 nm. The urease concentration was 0.2 μM.

Fluorescence correlation spectroscopy. For FCS measurement, enzyme (dye-labeled urease or acetylcholinesterase) were mixed with substrate in 100 mM PBS buffer (pH 7.2). To this, urea (Sigma) and 2 nM dye-labeled urease were added at room temperature. When studying acetylcholinesterase (AChE) reaction, 2 nM dye-labeled AChE were added in acetylthiocholine (Sigma) at room temperature. For hexokinase reaction, 2 nM of dye-labeled hexokinase I was added in the medium contained 50 mM Tris:HCl pH 7.5, 0.5 mM MgCl$_2$, 0.12 mM ATP, 0.1 mM NAD(P)+, and various concentrations of glucose as desired. FCS measurements were performed using a Leica TCS SP8X, using a 100× oil immersion objective lens with numerical aperture N.A. = 1.4 and pinhole size equal to 1 airy unit. Emitted fluorescence was collected using an avalanche photodiode (APD) (Micro Photon Devices; PicoQuant) through a 500- to 550-nm bandpass filter. The excitation power was controlled up to 20 μW. The APD signal was recorded using a time-correlated single–photon-counting (TCSPC) detection unit (Picoharp 300; PicoQuant).

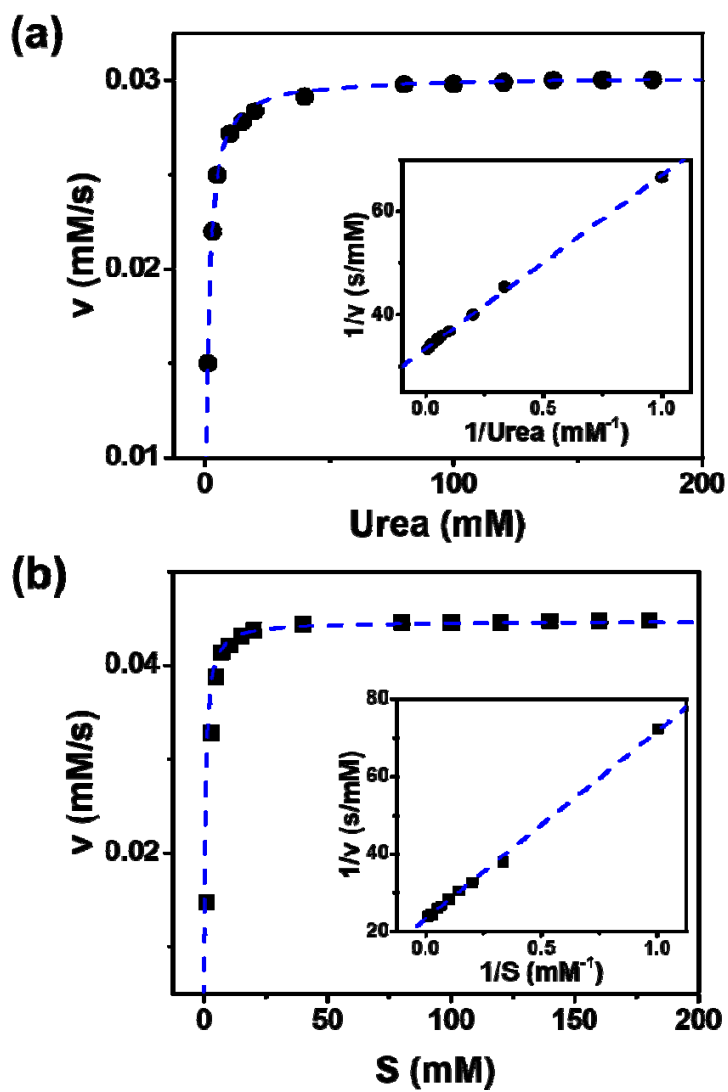

**Figure S1**. Assays of enzyme activity and fit to the Michaelis-Menten equation for urease sample Ur$_{1f}$ and acetylcholinesterase sample Ac$_f$. (a) Velocity of the urease reaction is plotted against urea concentration. Inset shows fit to the linearized form of the Michaelis-Menten equation. (b) Velocity of the acetylcholinesterase (AChE) reaction is plotted against acetylcholine concentration. Inset shows fit to the linearized form of the Michaelis-Menten equation. The fitted turnover rates of urease and AChE are 2,140 s$^{-1}$ and 4,500 s$^{-1}$, respectively. The fitted Michaelis-Menten constant ($k_M$) is 1.08 mM and 0.52 mM, respectively.

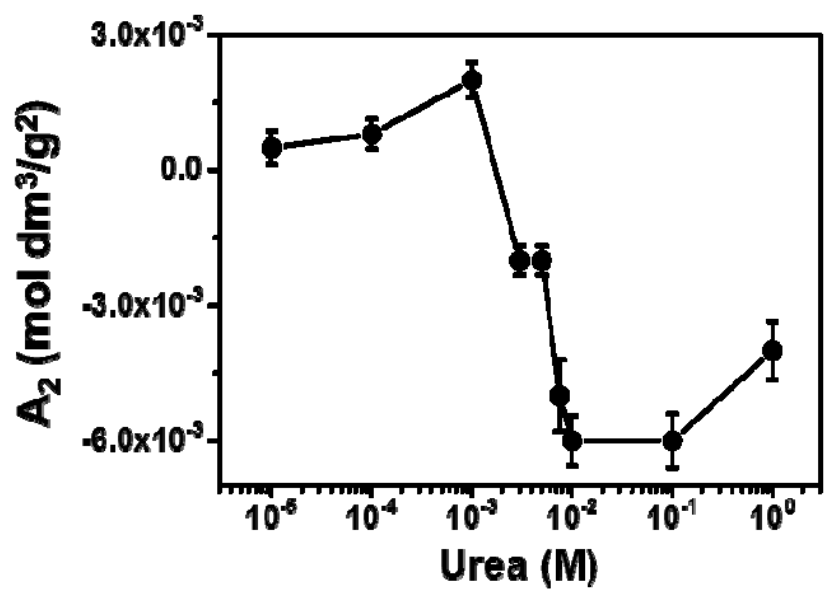

**Figure S2.** Second virial coefficient ($A_2$) as a function of substrate (urea) concentration, determined from static light scattering on urease sample $Ur_{1u}$. The negative second virial coefficient indicates attractive pairwise associations promoting aggregation.

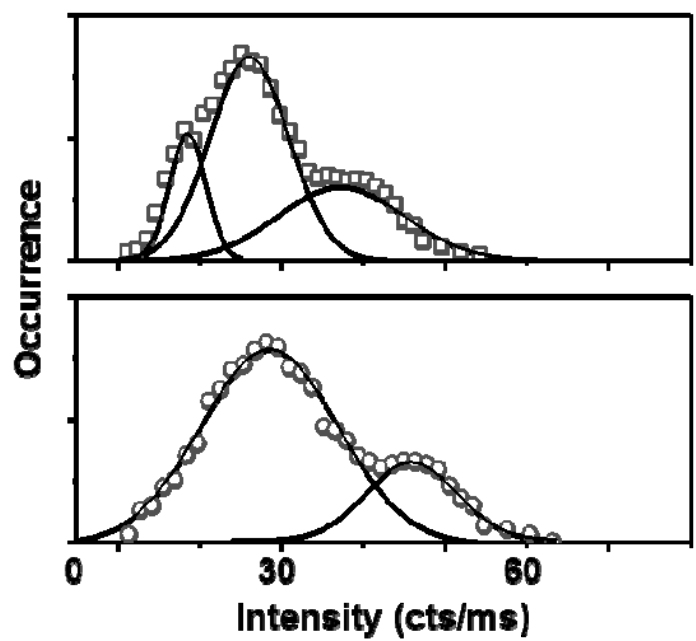

**Figure S3.** Fluorescence intensity distribution of urease sample $Ur_{1u}$ at 100 mM urea with 0.2 ms (upper panel) and 1 ms bin size (lower panel).

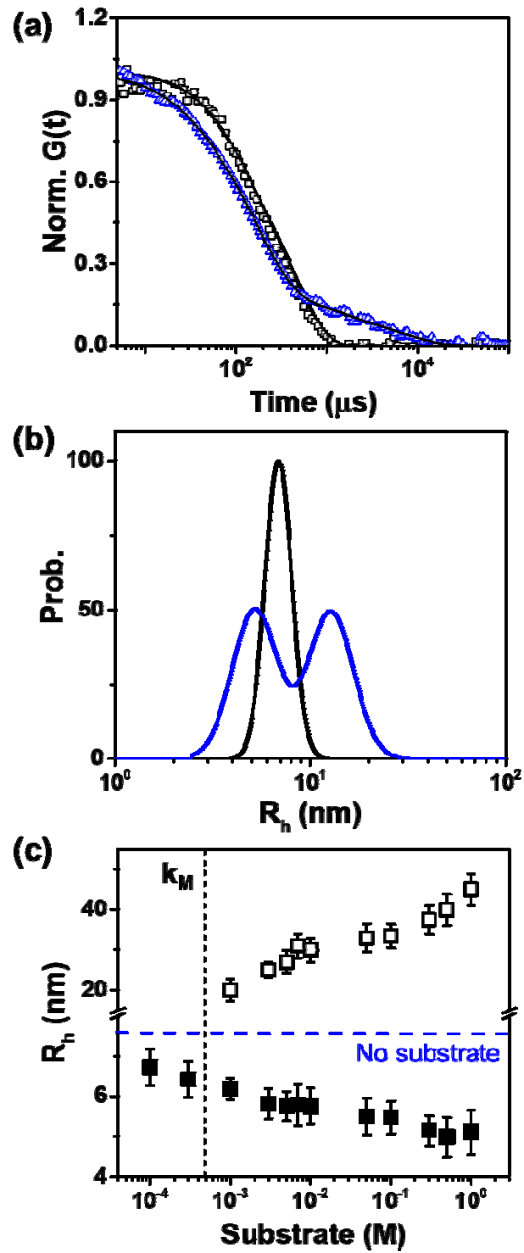

**Figure S.** Dynamic light scattering of dye-labeled acetylcholinesterase, sample Ac$_f$. (a) Photon autocorrelation function G(t) is plotted against time lag for 0.3 mM substrate (black) and 1 M substrate (blue). (b) Distribution of hydrodynamic radius $R_h$ inferred from the two curves in (a). (c) Hydrodynamic radius $R_h$ is plotted against substrate concentration. The measured $k_M$ is shown as a dotted vertical line. Above $k_M$, there is tendency to form aggregates (open symbols) and the smaller $R_h$ peak decreases indicating dissociation.

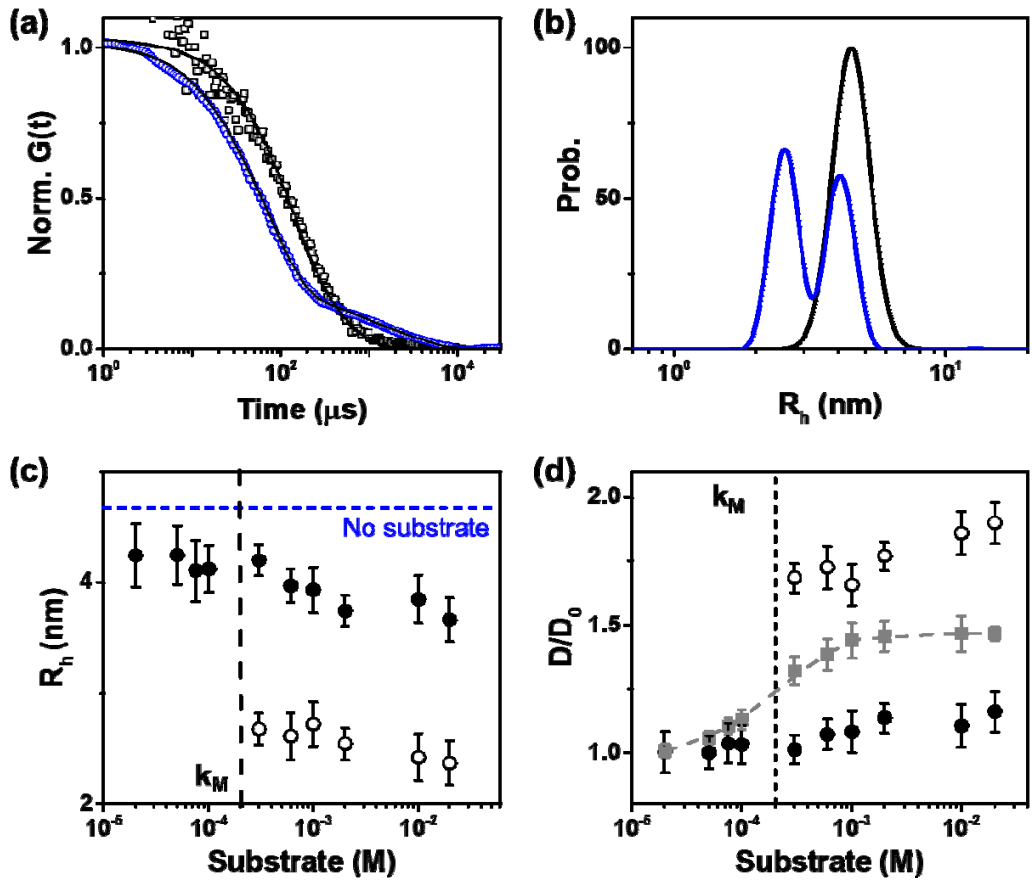

**Figure S5.** Dynamic light scattering of hexokinase, sample Hex. (a) Photon autocorrelation function G(t) is plotted against time lag for 0.02 mM substrate (black) and 20 mM substrate (blue). (b) Distribution of hydrodynamic radius $R_h$ inferred from data in (a). (c) Hydrodynamic radius $R_h$ is plotted against substrate concentration. The measured $k_M$ is shown as a dotted line. (d) The $R_h$ values of hexokinase for substrate obtained by DLS divided by $R_h$ with no substrate. Black filled symbol represents big size component, black empty symbols represent small size component. $D/D_0$ (grey squares) obtained by FCS is between the $D/D_0$ values of big and small components estimated from DLS.

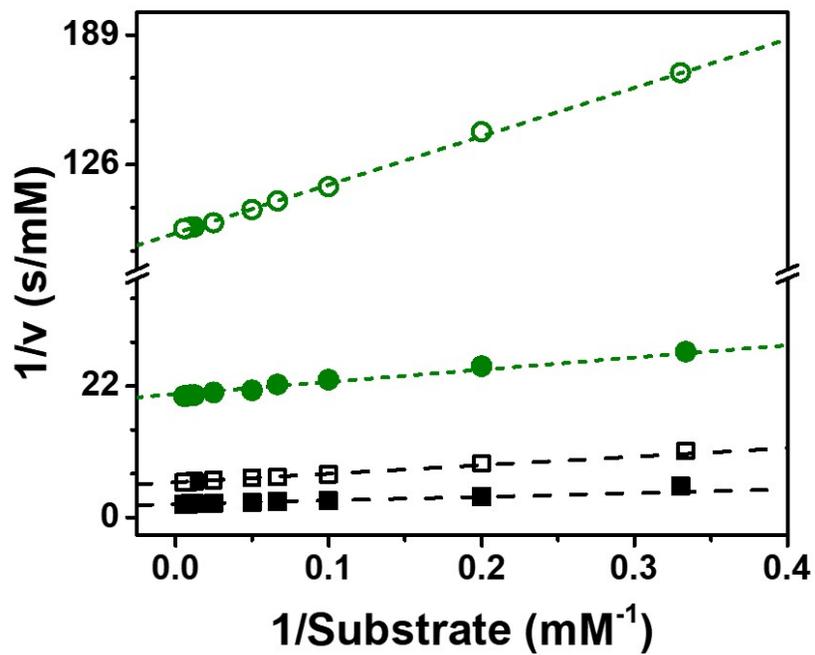

**Figure S6.** Assays of enzyme activity and fit to the linearized Michaelis-Menten equation for samples not shown in Fig. S1. Filled black and empty black shows $Ur_{2u}$ and $Ur_{2f}$ sample, respectively. Filled green and empty green shows $Ur_{1u}$ and $Ur_{1f}$ sample, respectively.

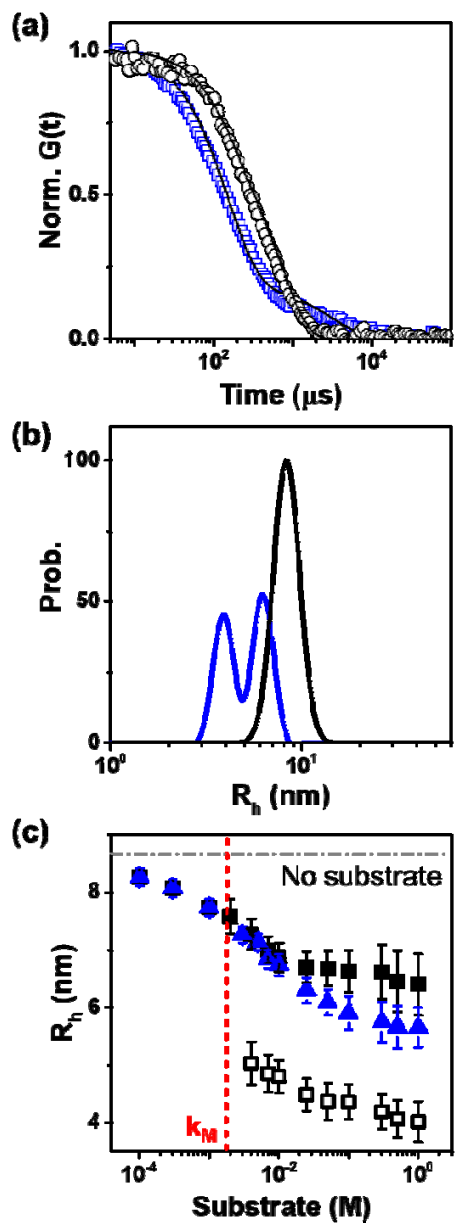

**Figure S7.** Dynamic light scattering of unlabeled urease, sample $Ur_{1u}$. (a) Photon autocorrelation function G(t) is plotted against time lag for 0.1 mM substrate (black) and 1 M substrate (blue). (b) Distribution of hydrodynamic radius $R_h$ inferred from the two curves in (a). (c) Hydrodynamic radius $R_h$ is plotted against substrate concentration. The measured $k_M$ is shown as a dotted vertical line. Above $k_M$, there is tendency to form aggregates (open symbols) and the smaller $R_h$ peak decreases indicating dissociation.

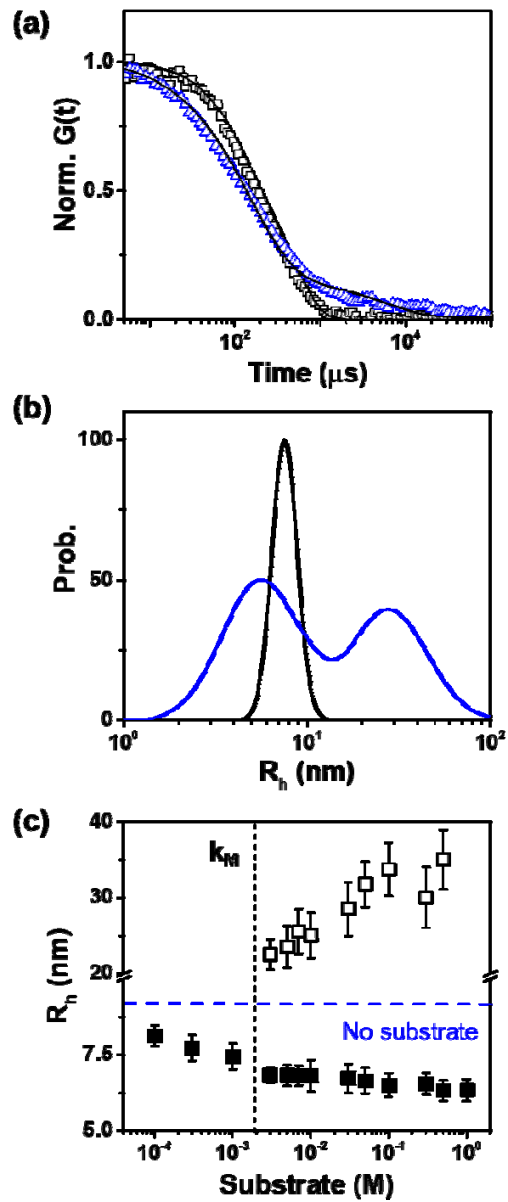

**Figure S8.** Dynamic light scattering of dye-labeled high-activity urease, sample $Ur_{2f}$. (a) Photon autocorrelation function G(t) is plotted against time lag for 0.1 mM substrate (black) and 1 M substrate (blue). (b) Distribution of hydrodynamic radius $R_h$ inferred from the two curves in (a). (c) Hydrodynamic radius $R_h$ is plotted against substrate concentration. The measured $k_M$ is shown as a dotted vertical line. Above $k_M$, there is tendency to form aggregates (open symbols) and the smaller $R_h$ peak decreases indicating dissociation.

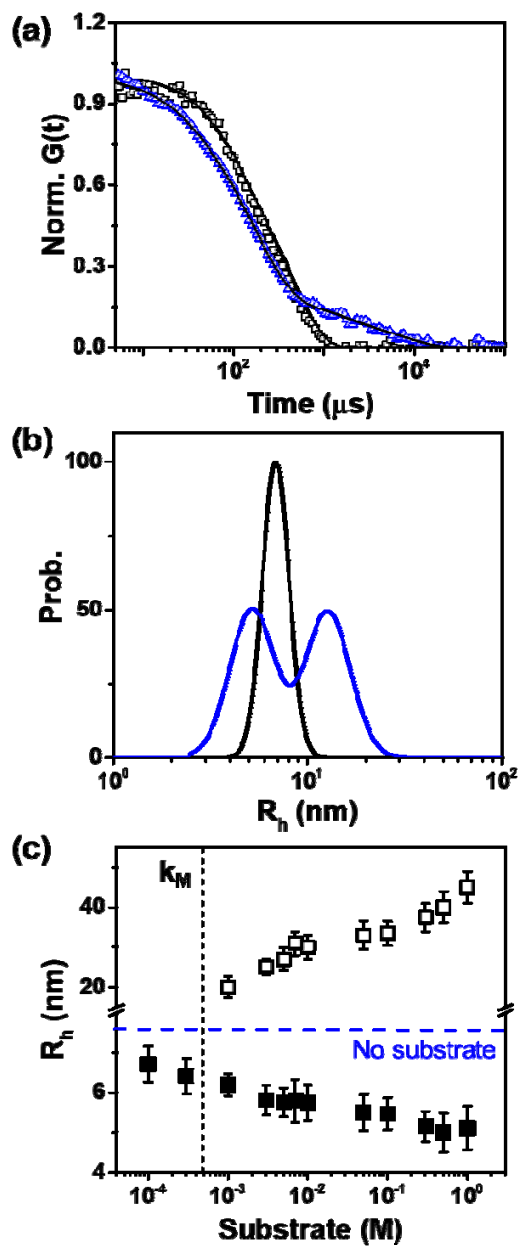

**Figure S9.** Dynamic light scattering of unlabeled acetylcholinesterase, sample Ac$_u$. (a) Photon autocorrelation function G(t) is plotted against time lag for 0.1 mM substrate (black) and 1 M substrate (blue). (b) Distribution of hydrodynamic radius $R_h$ inferred from the two curves in (a). (c) Hydrodynamic radius $R_h$ is plotted against substrate concentration. The measured $k_M$ is shown as a dotted vertical line. Above $k_M$, there is tendency to form aggregates (open symbols) and the smaller $R_h$ peak decreases indicating dissociation.

**Table S1.** Published literature on urease enhanced diffusion showing the reported source of urease in each study.

| Reference | Urease source | Catalog code | Substrate concentration range (mM) |
|---|---|---|---|
| 2 | Sigma | C-3 | 1~1000 |
| 3 | Sigma | C-3 | 1000 |
| 4 | Sigma | not mentioned | 1~1000 |
| 5 | Sigma | not mentioned | 500 |
| 6 | Sigma | not mentioned | 5 ~ 100 |
| 7 | TCI | not mentioned | 10 ~ 250 |
| 8 | Sigma | not mentioned | 1 |
| 9 | Sigma | C-3 | 1 |
| 10 | TCI | not mentioned | 100 |
| 11 | not mentioned | not mentioned | 1 ~25 |
| 12 | Sigma | Type IX | ~ 25 |
| 13 | Sigma | Type IX | 100 |
| 14 | not mentioned | not mentioned | 100 |
| 15 | Sigma | Type IX | 500 |
| 16 | Sigma | Type IX | ~ 100 |
| 17 | Sigma | Type IX | 100 |